\documentclass[%
 aip,
 amsmath,amssymb,
 reprint,%
]{revtex4-2}

\usepackage{color,soul}
\usepackage{amsmath}
\usepackage{overpic}
\usepackage{graphicx}% Include figure files
\usepackage{dcolumn}% Align table columns on decimal point
\usepackage{bm}% bold math
\usepackage[utf8]{inputenc}
\usepackage[T1]{fontenc}
\usepackage{mathptmx}
\usepackage{subfigure}

\makeatother

\begin{document}

\preprint{AIP/123-QED}

\title{Probing the Manipulation of Antiferromagnetic Order in CuMnAs Films Using Neutron Diffraction}

\author{S. F. Poole}
\affiliation{ 
School of Physics and Astronomy, University of Nottingham, University Park, Nottingham, NG7 2RD, United Kingdom
}

\author{L. X. Barton}
\affiliation{ 
School of Physics and Astronomy, University of Nottingham, University Park, Nottingham, NG7 2RD, United Kingdom
}

\author{M. Wang}
\affiliation{ 
School of Physics and Astronomy, University of Nottingham, University Park, Nottingham, NG7 2RD, United Kingdom
}

\author{P. Manuel}
\affiliation{
ISIS Facility, Rutherford Appleton Laboratory, Chilton, Didcot, OX11 0QX, United Kingdom
}

\author{D. Khalyavin}
\affiliation{
ISIS Facility, Rutherford Appleton Laboratory, Chilton, Didcot, OX11 0QX, United Kingdom
}

\author{S. Langridge}
\affiliation{
ISIS Facility, Rutherford Appleton Laboratory, Chilton, Didcot, OX11 0QX, United Kingdom
}

\author{K. W. Edmonds}
\affiliation{ 
School of Physics and Astronomy, University of Nottingham, University Park, Nottingham, NG7 2RD, United Kingdom
}

\author{R. P. Campion}
\affiliation{ 
School of Physics and Astronomy, University of Nottingham, University Park, Nottingham, NG7 2RD, United Kingdom
}

\author{V. Nov\'{a}k}
\affiliation{Institute of Physics, Czech Academy of Sciences, 162 00 Praha 6, Czech Republic}

\author{P. Wadley}
\email{Peter.Wadley@nottingham.ac.uk}
\affiliation{ 
School of Physics and Astronomy, University of Nottingham, University Park, Nottingham, NG7 2RD, United Kingdom
}

\date{\today}

\begin{abstract}
We describe measurements of the uniaxial magnetic anisotropy and spin-flop rotation of the N\'eel vector in antiferromagnetic CuMnAs thin films using neutron diffraction. The suppression of the magnetic (100) peak under magnetic fields is observed for films as thin as 20~nm indicating that they undergo a spin-flop transition. Good agreement is found between neutron diffraction and electron transport measurements of the spin-flop rotation in the same layer, with a similar shape and hysteresis of the obtained curves, while the neutron measurements provide a quantitative determination of the spin flop extent throughout the antiferromagnet layer. 
\end{abstract}

\maketitle

There is growing interest in antiferromagnetic (AF) materials for applications in spintronics. Predictions and demonstrations of current-induced switching of the AF N\'eel vector have stimulated the development of all-electrical AF memory devices, with potential for high speed operation, robustness against external fields, and multi-level neuromorphic outputs \cite{Zelezny2014, Wadley2016, Bodnar2018, Chen2018, Kaspar2021}. Electrical readout of the AF state typically utilizes anisotropic magnetoresistance (AMR) or spin Hall magnetoresistance, but such effects are small and can easily be obscured by structural changes induced by Joule heating \cite{Matalla2020}. Direct determination of the N\'eel vector is therefore important. This has been achieved by imaging local modifications of AF domains and domain walls affected by current pulses or magnetic fields using x-ray photoemission electron microscopy \cite{Grzybowski2017, Wadley2018, Sapoznik2018, Bodnar2019, Wang2020}, and more recently by optical birefringence \cite{Meer2021}, spin-Seebeck \cite{Gray2019} or magneto-Seebeck \cite{Janda2020} microscopies.

Neutron diffraction is a well-established technique to directly measure the atomic arrangement of magnetic moments in a bulk AF crystal. It provides a quantitative determination of the N\'eel vector and, unlike electron microscopy techniques, is routinely performed in high magnetic fields. It is therefore well-suited to studying spin-flop transitions, which typically occur in magnetic fields of several Tesla. In AF thin films and devices, spin-flop rotations of the N\'eel vector are invaluable for characterizing extraordinary magnetotransport \cite{Wang2020} or magnon transport phenomena \cite{Bender2017,Ross2020}, and exploring multi-stability of magnetic memory states \cite{Kriegner2016,Kriegner2017}. However, there are few direct studies of spin-flop rotation in AF thin films using neutron diffraction due to the very small volume of the sample film compared to a typical neutron experiment, which is a flux limited technique.

Here, we demonstrate that neutron diffraction can be used in an unambiguous total volume determination of the spin-flop rotation in CuMnAs films with thickness down to 20~nm. We also show good agreement between neutron diffraction and magnetotransport measurements performed on the same sample. CuMnAs is a collinear antiferromagnet with a N\'eel temperature of around 480~K \cite{Wadley2015}. It can exhibit pronounced modifications of its AF domain structure in response to applied electrical or optical pulses, which may be accompanied by significant electrical readout signals \cite{Kaspar2021}. Understanding of its magnetic and transport properties is therefore important for the design of spintronic memory devices. 

% The CuMnAs 20~nm and 45~nm films were grown on GaP(001) substrates using molecular beam epitaxy \cite{krizek2020molecular}. X-ray diffraction measurements confirmed the tetragonal crystal structure of the films, shown in Fig. \ref{fig:ISIS_schemes}(a), with the [001] axis perpendicular to the plane and the [100] axis parallel to the substrate [110]. Previous neutron diffraction and x-ray magnetic linear dichroism measurements on similar films have shown that the magnetic moments align in the $ab$ plane, with a competition between a uniaxial [100] anisotropy and a biaxial [110]/[1$\bar{1}$0] anisotropy \cite{Wadley2015}.

\begin{figure}[t]
    \centering
    \includegraphics[width=0.45\textwidth]{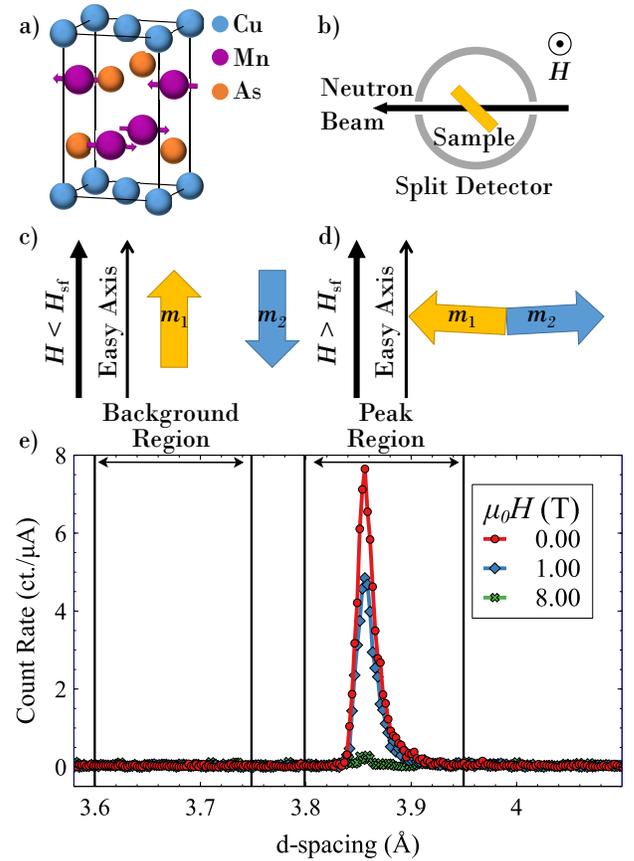}
    \caption{\label{fig:ISIS_schemes} a) CuMnAs unit cell with arrows representing the expected uniaxial anisotropy along the [010] crystal direction. (b) Schematic diagram of the neutron diffraction experiment showing the horizontal scattering plane and the vertical magnetic field. The incident beam is represented by the black arrow, and the detector arrays covering in-plane scattering angles from 10$^\circ$ to 170$^\circ$ are depicted as gray semi-circles. (c,d) Sketches showing the magnetic moment orientations for an uniaxial antiferromagnet with respect to an applied magnetic field below (c) and above (d) the spin-flop field $H_{sf}$. e) Plots of the (100) peak for the 45~nm film with three different field strengths applied along the easy axis. The integration regions used in background subtraction are shown.}
\end{figure}

The CuMnAs 20~nm and 45~nm films were grown on GaP(001) substrates using molecular beam epitaxy \cite{krizek2020molecular}. X-ray diffraction measurements confirmed the tetragonal crystal structure of the films, illustrated in Fig. \ref{fig:ISIS_schemes}(a), with the CuMnAs[001] axis perpendicular to the plane and the CuMnAs principal axes at $45^\circ$ to the substrate principal axes. Previous neutron diffraction and x-ray magnetic linear dichroism measurements on similar films have shown that the magnetic moments align in the $ab$ plane, with a competition between a uniaxial CuMnAs[010] anisotropy and a biaxial CuMnAs[110]/[1$\bar{1}$0] anisotropy \cite{Wadley2015}. The uniaxial in-plane anisotropy arises because of the substrate-film interface where the symmetry is broken by the P-terminated unit cell of the GaP substrate.

The neutron diffraction measurements utilized the time of flight cold neutron beam line WISH at the ISIS facility\cite{chapon2011wish}. Figure \ref{fig:ISIS_schemes}(b) represents the geometry of the experiment with horizontal scattering plane and vertical magnetic field generated by a superconducting split-pair solenoid magnet. For both films, the samples were mounted in two orientations: in one case, the uniaxial easy axis ([010]) was parallel to the applied magnetic field, and in the other the field was parallel to the [100] axis. These corresponded to the (h0l) and (0kl) scattering plane, respectively. The position of the samples in respect of the incident beam was chosen to optimize the neutron flux for the (100) and (010) magnetic reflections as well as to access the (001) reflection at the same time. It was achieved when the surface plane of the samples was at $45^\circ$ to the beam direction. The measurements were conducted at a sample temperature of 100~K and the sample position was the same for both films investigated.

% The neutron diffraction measurements utilized the WISH detector at the ISIS facility. Figure \ref{fig:ISIS_schemes}(b) represents the geometry of the experiment with the sample placed inside a cryostat containing a superconducting solenoid magnet. For both films, the samples were mounted in two orientations: in one case, the uniaxial easy axis (CuMnAs[100] direction) was parallel to the applied magnetic field, and in the other the field was parallel to the CuMnAs[010] direction. In both orientations, the surface plane was at $45^\circ$ to the incident neutron beam, rotated around the field axis. The two $170^\circ$ coverage detectors simultaneously measured peak locations and time of flight information of the scattered neutrons. The measurements were conducted at a sample temperature of 100K.

For a magnetic field applied parallel to the easy axis of a uniaxial antiferromagnet (Fig. \ref{fig:ISIS_schemes}(c)), there is a threshold field strength known as the spin flop field, $H_\textup{sf}$, that causes a 90$^\circ$ reorientation of the antiferromagnetic N\'eel vector. Above this field, the moments are canted around the external field direction (Fig. \ref{fig:ISIS_schemes}(d)). Due to higher-order magnetic anisotropies as well as domain walls and other inhomogeneities, the sharp transition at $H_\textup{sf}$ is typically broadened into an s-shaped curve.

Nuclear contribution to the (100) and (010) reflections is forbidden in tetragonal CuMnAs (P4/nmm space group) and therefore they contain only magnetic scattering which is sensitive to the component of magnetic moments perpendicular to the scattering vector. Consider the case where the [010] direction is a uniaxial easy magnetic axis and the field is to be applied along this direction. In the zero field state where $H < H_\textup{sf}$, the magnetic moments are parallel to the easy axis and the (100) peak is observable. Under a field $H > H_\textup{sf}$, the moments are perpendicular to the easy axis and aligned along the [100] direction with small-angle canting of the moments in the direction of the field and therefore the (100) peak is extinguished. Hence, the magnitude of the peak provides a direct measurement of the spin flop transition in the sample. If only a partial rotation of the magnetic moments occurs, the magnitude of the peak will reduce in proportion, so the ratio of the peak amplitude at field compared to the original state will give a direct measurement of the extent of spin flop in the sample.

\begin{figure}[t]
    \centering
    \includegraphics[width=0.45\textwidth]{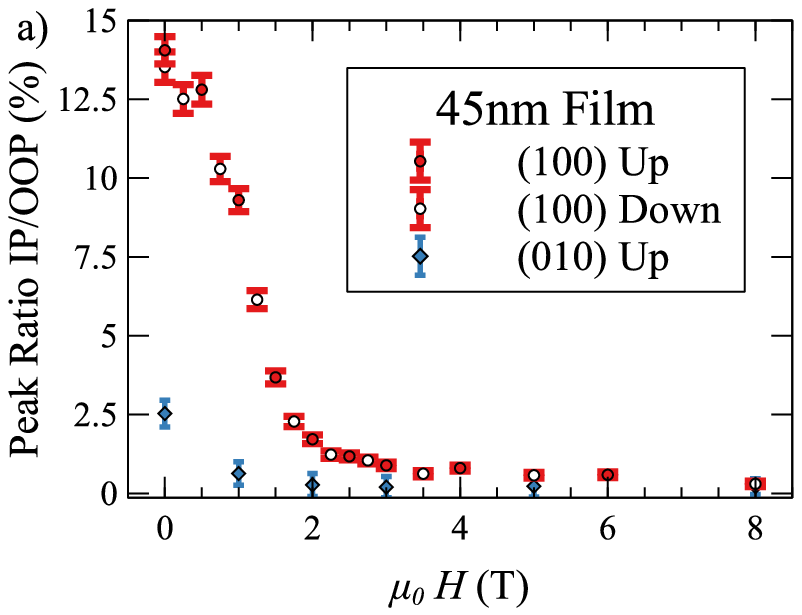}
    \includegraphics[width=0.45\textwidth]{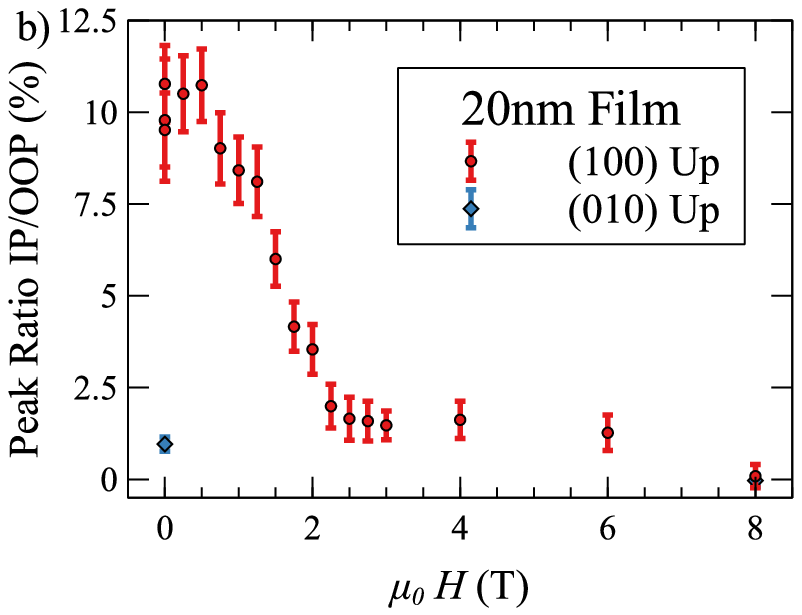}
    \caption{\label{fig:scattering_results} Ratio of in plane (IP) with the sample surface to out-of-plane (OOP) integrated peak amplitudes versus magnetic field, for CuMnAs films of thickness (a) 45~nm and (b) 20~nm. The red circles and blue diamonds correspond to to the measurements of the (100) and (010) in-plane peaks, with the field along the [010] and [100] axes respectively. Open and filled symbols represent decreasing and increasing field sweeps.}
\end{figure}

The (100) peak for a 45~nm CuMnAs film is shown in Fig. \ref{fig:ISIS_schemes}(e) for three different magnetic field strengths. The peak is significantly reduced at 1~T applied field compared to its zero-field amplitude, and has almost vanished at 8~T. To quantify the spin-flop transition, the measured intensity is integrated across the peak and background regions indicated in Fig. \ref{fig:ISIS_schemes}(e). The combined magnetic and structural (001) peak is also measured, to allow calibration of the alignment and detection efficiency. Because the moments are expected to stay within the (001) plane, the (001) peak is unaffected by spin flop and is used as a control measurement. Consistent with this, the peak amplitude is found to be independent of the magnetic field within the experiment uncertainty (see Supplementary Information).

Plots of the ratios of integrated peak amplitudes versus magnetic field for 45~nm and 20~nm thick CuMnAs films are shown in Fig.~\ref{fig:scattering_results}. For the 45~nm sample (Fig.~\ref{fig:scattering_results}a), the (100) peak is larger than the (010) peak by a factor of $5.4 \pm 0.9$. This indicates that the magnetic moments in the sample are predominantly aligned with the [010] axis. A sharp decrease in intensity is observed between zero and 1~T for the (010) peak, and between zero and 2~T for the (100) peak, due to the spin flop rotation of the N\'eel vector into an axis which is perpendicular to the field. In high fields, the (100) peak is decreased to $2.2 \pm 0.7$~\% of its zero-field amplitude, showing conclusively that the vast majority of the CuMnAs layer undergoes spin flop in this field range. After the high field measurements the sample recovers its original state. This observation in combination with the well-defined critical fields strongly suggests that the reorientation of the magnetic moments is a flop-transition rather than due to continuous moving domain walls.

The 20~nm CuMnAs layer (Fig.~\ref{fig:scattering_results}b), exhibited a (100) to (010) peak amplitude ratio of $10.5 \pm 3.4$ and therefore demonstrates even stronger anisotropy than the thicker layer. As with the thicker layer, the (100) peak intensity drops sharply between 1~T and 2~T applied field. However, there is a significant intensity remaining between 2~T and 6~T, which may be attributed to the presence of $180^\circ$ magnetic domain walls and defect-driven domain pinning. The intensity then drops to zero between 6~T and 8~T. Due to time constraints and the weak signal from this thin film, the (010) peak was not measured over the whole magnetic field range.

%It is typical for thicker ($>$30~nm) films to exhibit biaxial [110]/[1$\bar{1}$0] anisotropy and for thin films, such as 20~nm, to exhibit uniaxial [100] anisotropy\cite{Wadley2015, Wang2020}. The thick film exhibiting uniaxial behaviour may be due to being grown on a sulphur doped substrate which may extend the strain induced magnetic anisotropy throughout the film.

\begin{figure}[t]
     \centering
     \includegraphics[width=0.4\textwidth]{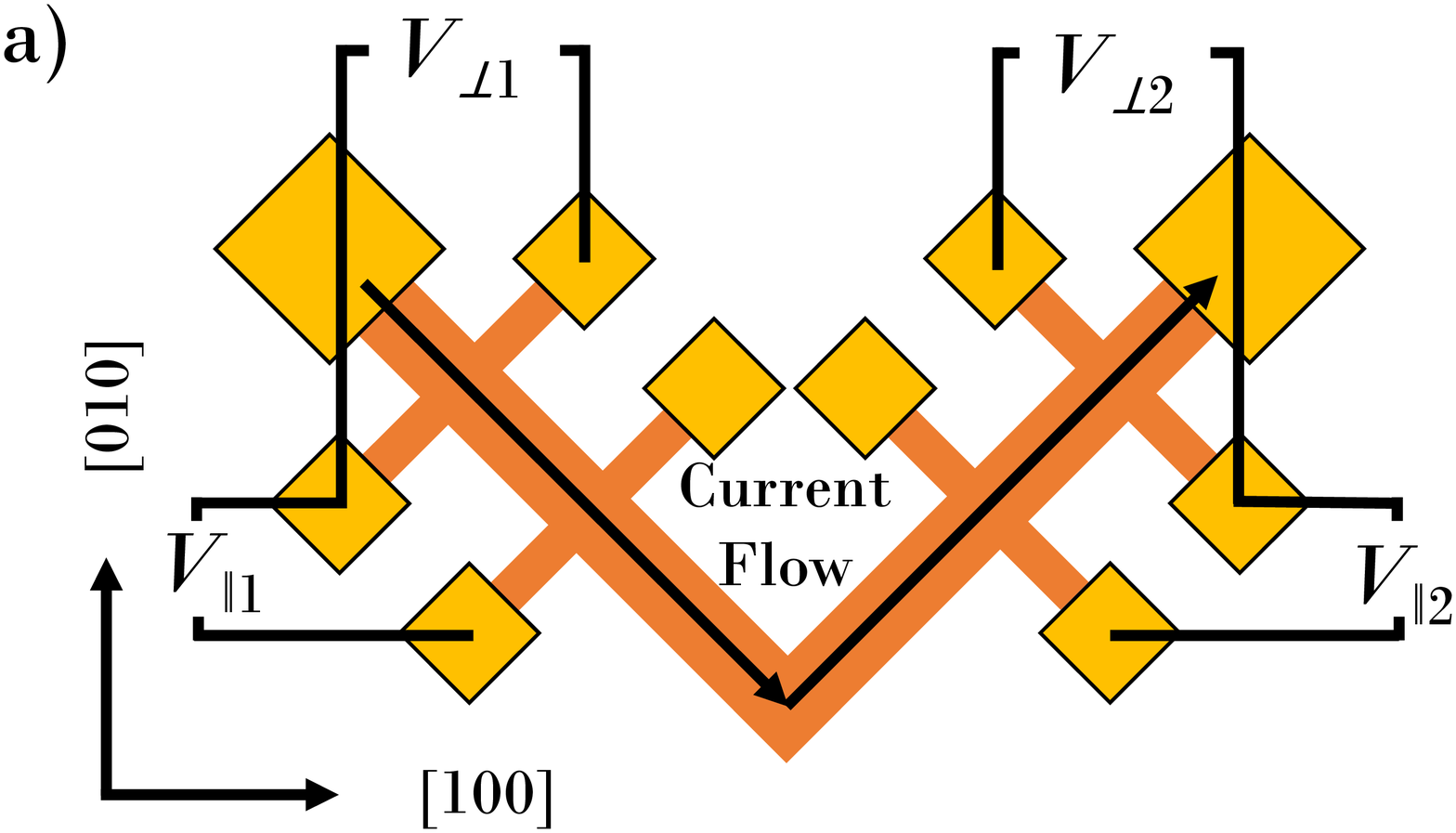}
     \includegraphics[width=0.4\textwidth]{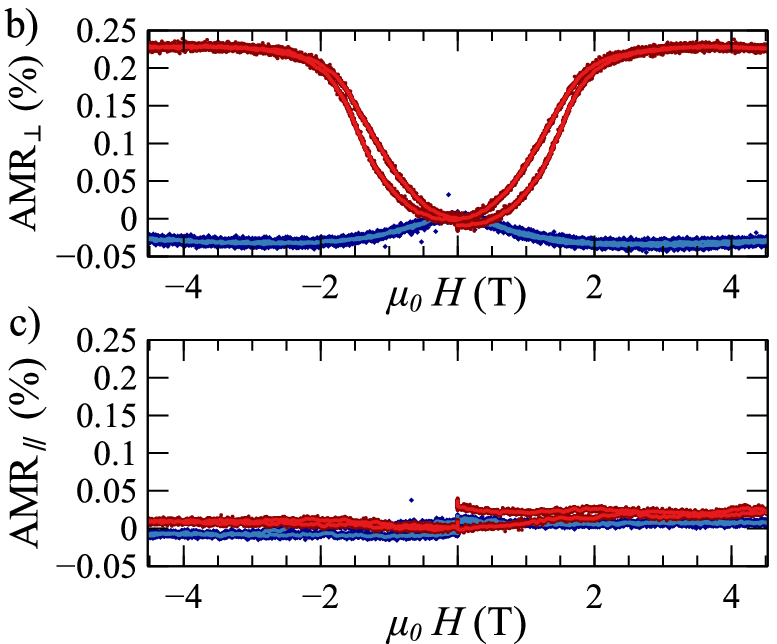}
    \caption{\label{fig:AMR} (a) Schematic of an L-bar device. The orange represents the 100~{\textmu}m wide CuMnAs bar and the gold squares represent contact pads to which wires are bonded. The arrows represent the current flow in the device and the voltages drawn correspond to the voltages described in the text where 1 and 2 correspond to the first and second arm the current flows through respectively. (b) Transverse AMR and (c) longitudinal AMR, as defined by Eq. 1 and Eq. 2 respectively, versus magnetic field for an L-bar device fabricated from the 45~nm thick CuMnAs film. Red and blues lines are for field along the [010] and [100] axes, respectively. The plots are offset for clarity to show zero AMR at zero field.}
\end{figure}

\begin{figure}[t]
    \centering
    \includegraphics[width=0.3\textwidth]{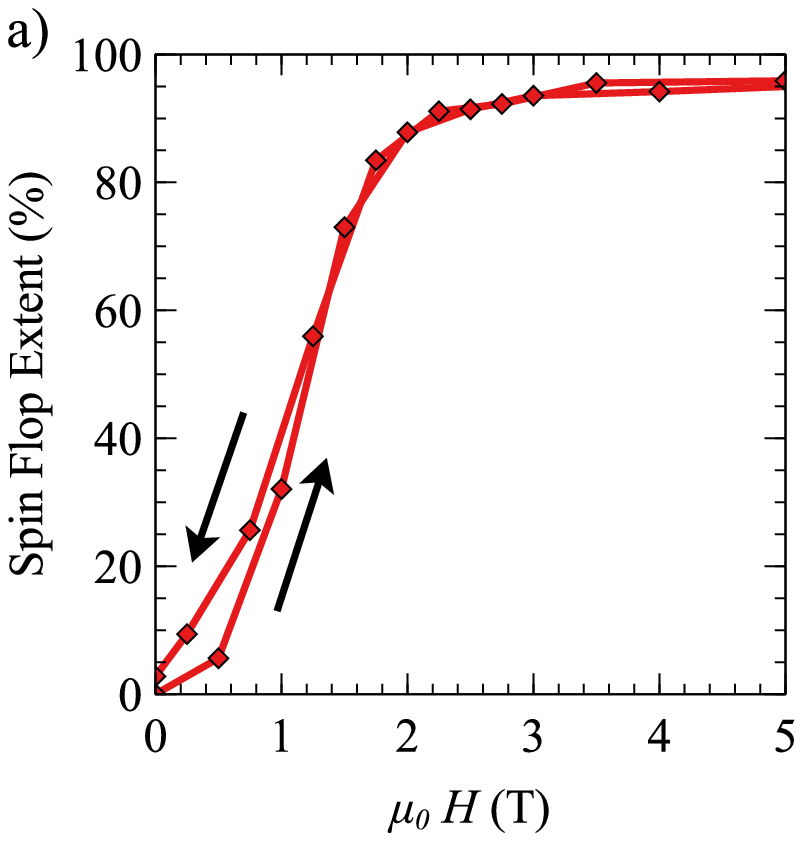}
    \includegraphics[width=0.3\textwidth]{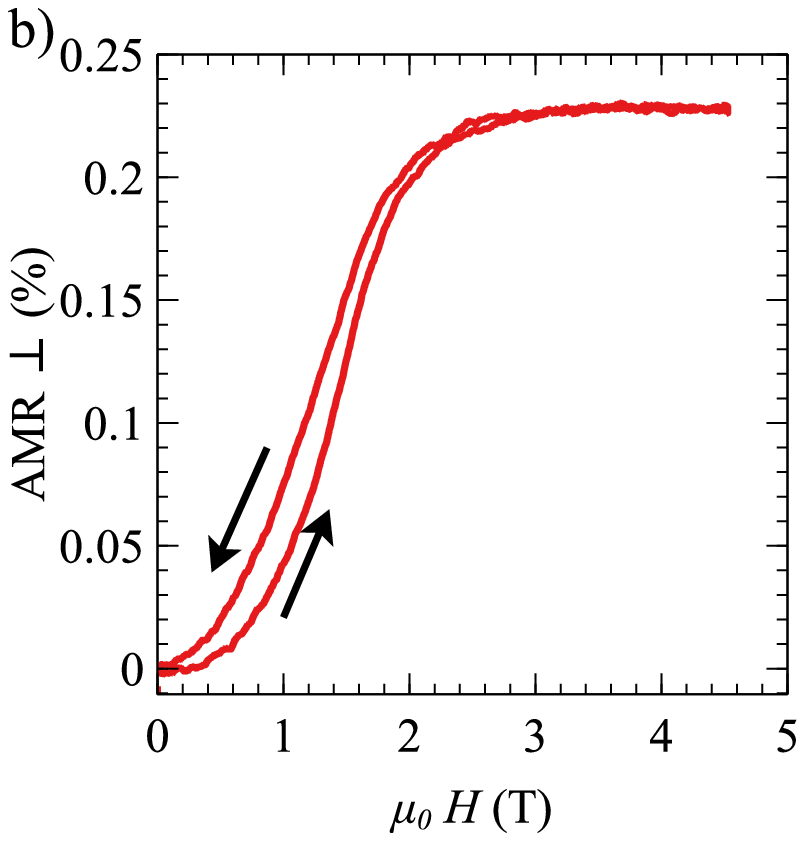}
    \caption{\label{fig:Comparison} (a) Spin flop extent extracted from neutron diffraction, and b) transverse AMR, for the 45nm thick CuMnAs film with external magnetic field applied along the [010] axis. Arrows indicate the direction of the magnetic field sweep.}
\end{figure}

Complementary electrical transport measurements were conducted on the 45~nm sample which was patterned into an L-shaped Hall bar structure as illustrated in Fig.~\ref{fig:AMR}a. This allows currents to be applied simultaneously along two in-plane crystal axes, at $45^\circ$ and $-45^\circ$ to an applied magnetic field respectively, while recording the longitudinal and transverse voltages $V_\parallel$ and $V_{\perp}$. Each arm of the structure is of width $\bm{100}\mu$m. The magnetoresistance measurements were conducted with the sample submerged in liquid helium at 4K inside a superconducting magnet. Magnetic fields were applied either parallel to [010] or [100], with applied current $I_{probe}=2$~mA. 

Due to anisotropic magnetoresistance (AMR), the longitudinal and transverse voltages depend on the angle $\phi$ between the applied current and the orientation of the N\'eel vector. Neglecting crystalline terms, this dependence is given by $V_\parallel=V_0 + V_1\cos(2\phi)$, and $V_\perp=V_1\sin(2\phi)$. Hence, the rotation of the N\'eel vector from $+45^\circ$ to $-45^\circ$ with respect to the current direction should generate no change in the longitudinal voltages, and a maximum change in the transverse voltages. 

Figures~\ref{fig:AMR}b and \ref{fig:AMR}c show the measured normalized transverse and longitudinal anisotropic magnetoresistances, defined as
\begin{align}
     AMR_\perp &= \frac{2(V_{\perp 1} - V_{\perp 2})}{V_{\parallel 1} + V_{\parallel 2}} \\
    AMR_\parallel &= \frac{V_{\parallel 1} - V_{\parallel 2}}{V_{\parallel 1} + V_{\parallel 2}}    
\end{align}

\noindent where the factor of 2 for $AMR_\perp$ accounts for the number of squares between the longitudinal voltage contacts. For both field directions, a broad s-shaped change in $AMR_\perp$ is observed below 2~T. A hysteresis is observed between the increasing and decreasing field sweeps, but the values return to the same point at zero field. The variations in $AMR_\parallel$ are much smaller and are ascribed to small variations in the sample temperature during the field sweep.

Figure \ref{fig:Comparison} compares the neutron diffraction and $AMR_\perp$ measurements over the same magnetic field range. The spin flop extent is defined as $ 1 - \frac{P(H)}{P(0)}$ where $P(H)$ is the integrated peak amplitude at field $H$ and ${P(0)}$ is the initial value at zero field. It can be seen that the transport measurement gives a qualitatively accurate measure of the spin flop extent, with the shape, hysteresis and spin flop field all matching well. Note that the neutron diffraction measurements were performed at 100~K while the AMR was measured at 4~K. Previous studies have shown that this difference in temperature creates a small change in amplitude of the AMR signal due to changes in base resistance, but does not qualitatively change the signal \cite{Wang2020}.

We have demonstrated the use of magnetic neutron scattering as a tool for probing the antiferromagnetic spin-flop regime in CuMnAs. From our results we were able to show that 98\% of the magnetic order is reoriented under action of an applied magnetic field along the magnetic easy axis in a strongly uniaxial sample. Any remnant signal could be attributed to $180^\circ$ domain walls and domain pinning due to defects. Complementary electrical transport measurements were congruent with the neutron scattering signal in both shape and hysteretic behaviour. Overall, we propose that this technique should be applicable for measuring the magnetic N\'eel vector magnitude and direction of any such compatible magnetic crystals.

\section*{Supplementary Material}
See supplementary material for plots of the peak amplitudes that were used in the normalization process to make Fig.~\ref{fig:scattering_results}. The supplementary material also includes waterfall plots of the diffraction peaks used to find the amplitudes in the aforementioned plots.

\begin{acknowledgments}
The work was supported by the EU FET Open RIA Grant no 766566 and the UK Engineering and Physical Sciences Research Council [grant number EP/P019749/1]. We wish to acknowledge ISIS Neutron and Muon Source for the beamtime on WISH. Data is available here: \url{https://doi.org/10.5286/ISIS.E.RB1920313}
\end{acknowledgments}

\bibliography{Neutrons}

\end{document}

\preprint{AIP/123-QED}

\title{Supplementary Information for Probing the Manipulation of Antiferromagnetic Order in CuMnAs Films Using Neutron Diffraction}

\author{S. F. Poole}
\affiliation{ 
School of Physics and Astronomy, University of Nottingham, University Park, Nottingham, NG7 2RD, United Kingdom
}

\author{L. X. Barton}
\affiliation{ 
School of Physics and Astronomy, University of Nottingham, University Park, Nottingham, NG7 2RD, United Kingdom
}

\author{M. Wang}
\affiliation{ 
School of Physics and Astronomy, University of Nottingham, University Park, Nottingham, NG7 2RD, United Kingdom
}

\author{P. Manuel}
\affiliation{
ISIS Facility, Rutherford Appleton Laboratory, Chilton, Didcot, OX11 0QX, United Kingdom
}

\author{D. Khalyavin}
\affiliation{
ISIS Facility, Rutherford Appleton Laboratory, Chilton, Didcot, OX11 0QX, United Kingdom
}

\author{S. Langridge}
\affiliation{
ISIS Facility, Rutherford Appleton Laboratory, Chilton, Didcot, OX11 0QX, United Kingdom
}

\author{K. W. Edmonds}
\affiliation{ 
School of Physics and Astronomy, University of Nottingham, University Park, Nottingham, NG7 2RD, United Kingdom
}
\email{Kevin.Edmonds@nottingham.ac.uk}

\author{R. P. Campion}
\affiliation{ 
School of Physics and Astronomy, University of Nottingham, University Park, Nottingham, NG7 2RD, United Kingdom
}

\author{V. Nov\'{a}k}
\affiliation{Institute of Physics, Czech Academy of Sciences, 162 00 Praha 6, Czech Republic}

\author{P. Wadley}
\email{Peter.Wadley@nottingham.ac.uk}
\affiliation{ 
School of Physics and Astronomy, University of Nottingham, University Park, Nottingham, NG7 2RD, United Kingdom
}

\date{\today}

\maketitle
\supplementarysection

% \section{Additional Figures}

\begin{figure}
    \centering
    \includegraphics[width=0.45\textwidth]{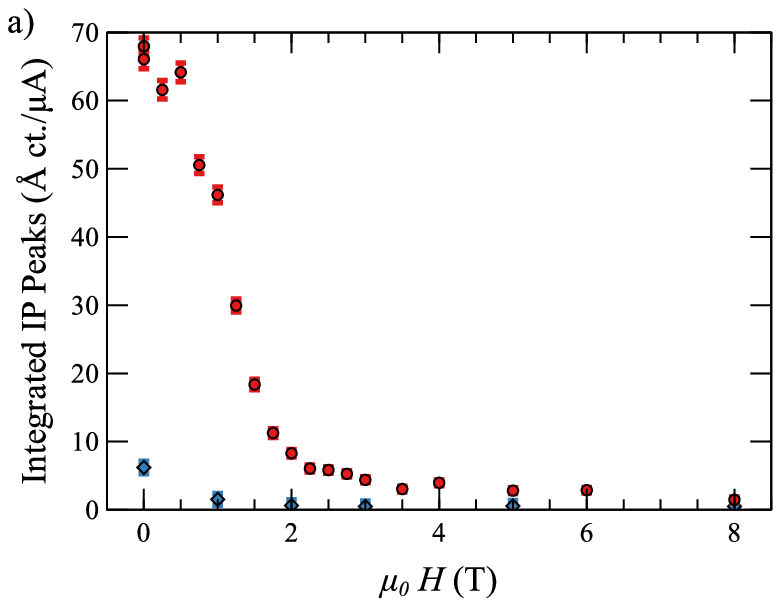}
    \includegraphics[width=0.45\textwidth]{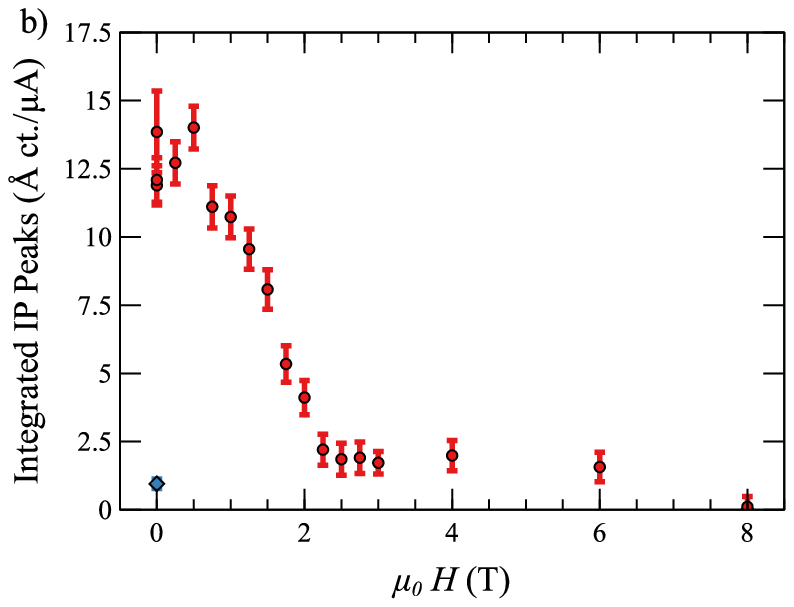}\\
    \includegraphics[width=0.45\textwidth]{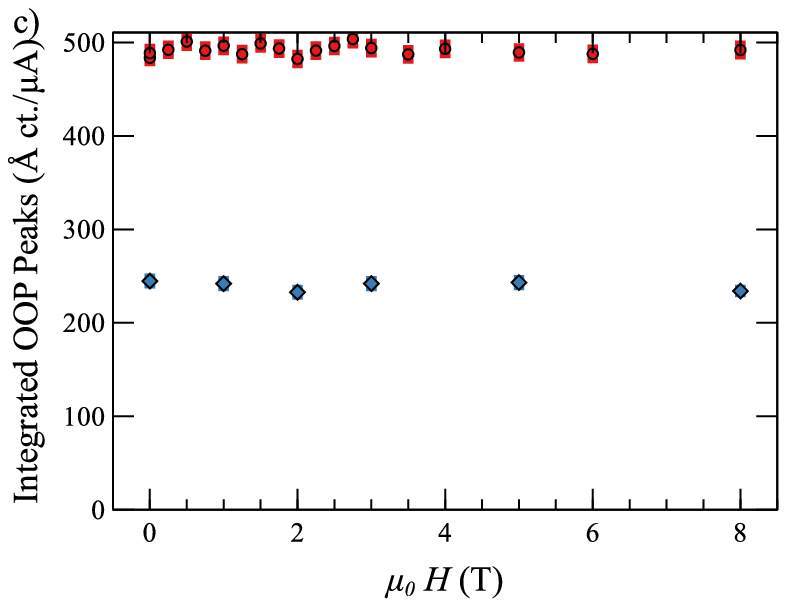}
    \includegraphics[width=0.45\textwidth]{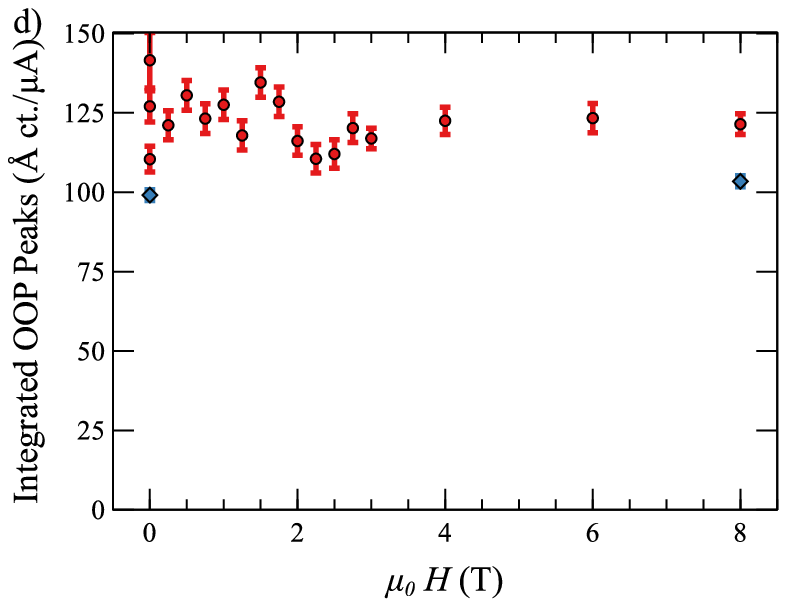}
    \caption{\label{fig:RC194_supp_scattering} Neutron diffraction peak amplitudes for the 45nm thick film (a,c) and 20nm thick film (b,d) comprised from the data in Figs. S2 to S5. The top row (a,b) correspond to the purely magnetic in plane (IP) peaks and the bottom row correspond to the structural magnetic out of plane (OOP) peaks. The red dots correspond to measurements with the magnetic field applied along the easy axis and the blue diamonds correspond to the measurements with the magnetic field applied along the hard axis. The values plotted in Figure 2 of the main text were calculated by dividing the points in the top row figures by the corresponding values in the bottom row figures.}
\end{figure}

\begin{figure}
    \centering
    \includegraphics[width=0.45\textwidth]{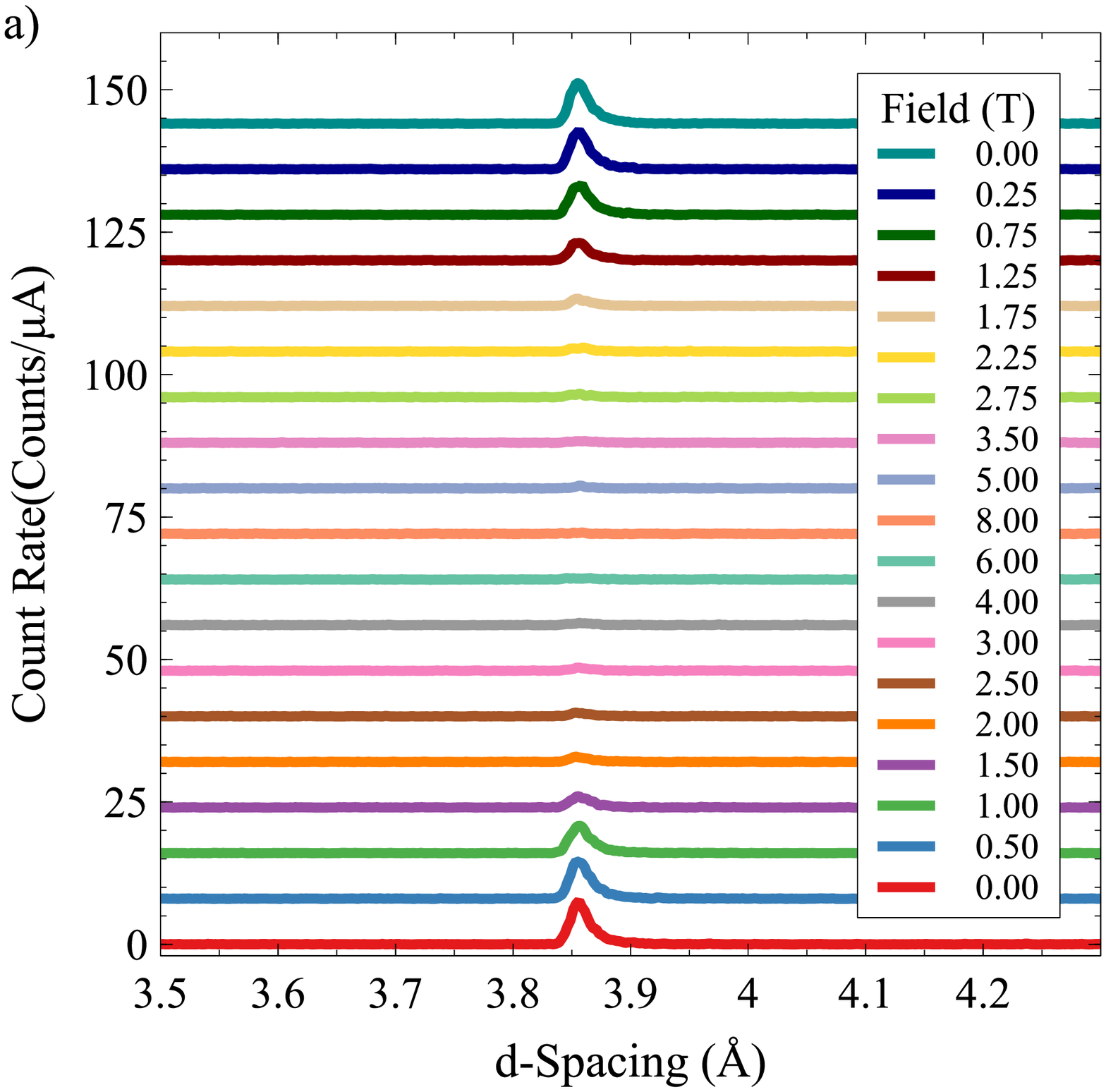}
    \includegraphics[width=0.454\textwidth]{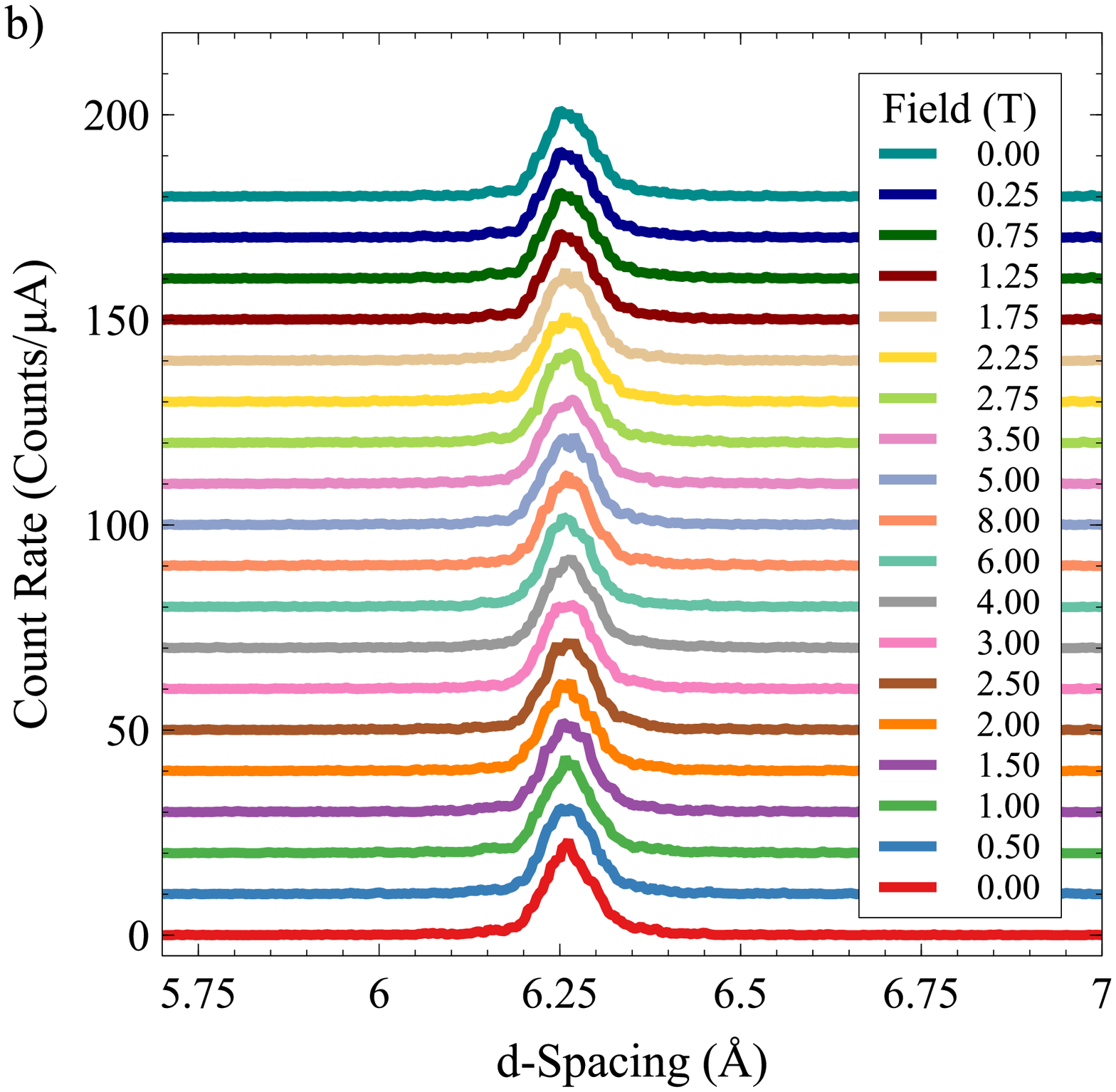}
    \caption{\label{fig:rc194bwaterfalls} Waterfall plots of  a) in plane and b) out of plane peaks for the 45~nm film with increasing field applied along the easy axis. The peak amplitudes correspond to the red circles in Figure S1 a,c.}
\end{figure}

\begin{figure}
\centering
\includegraphics[width=0.45\textwidth]{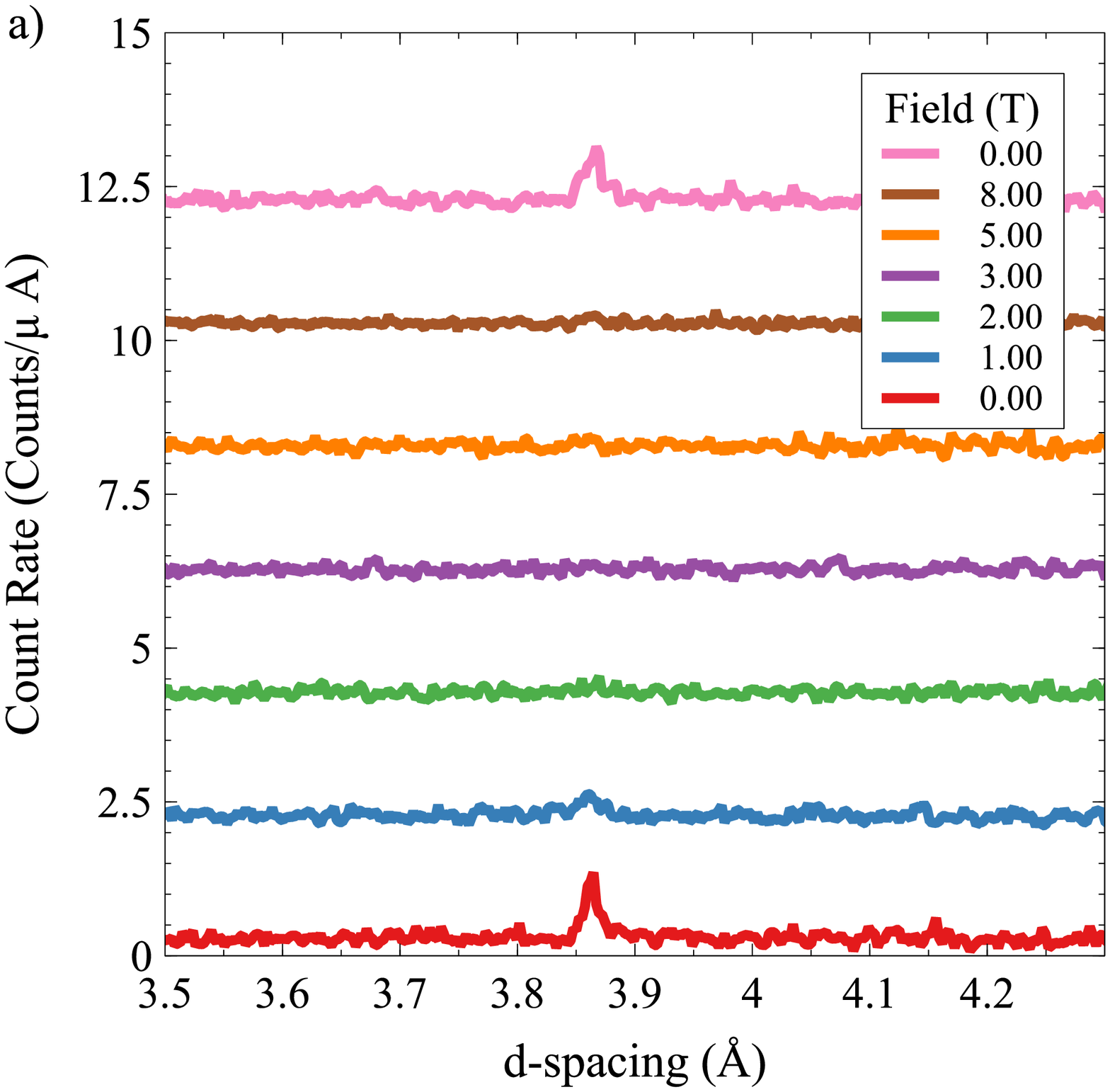}
\includegraphics[width=0.454\textwidth]{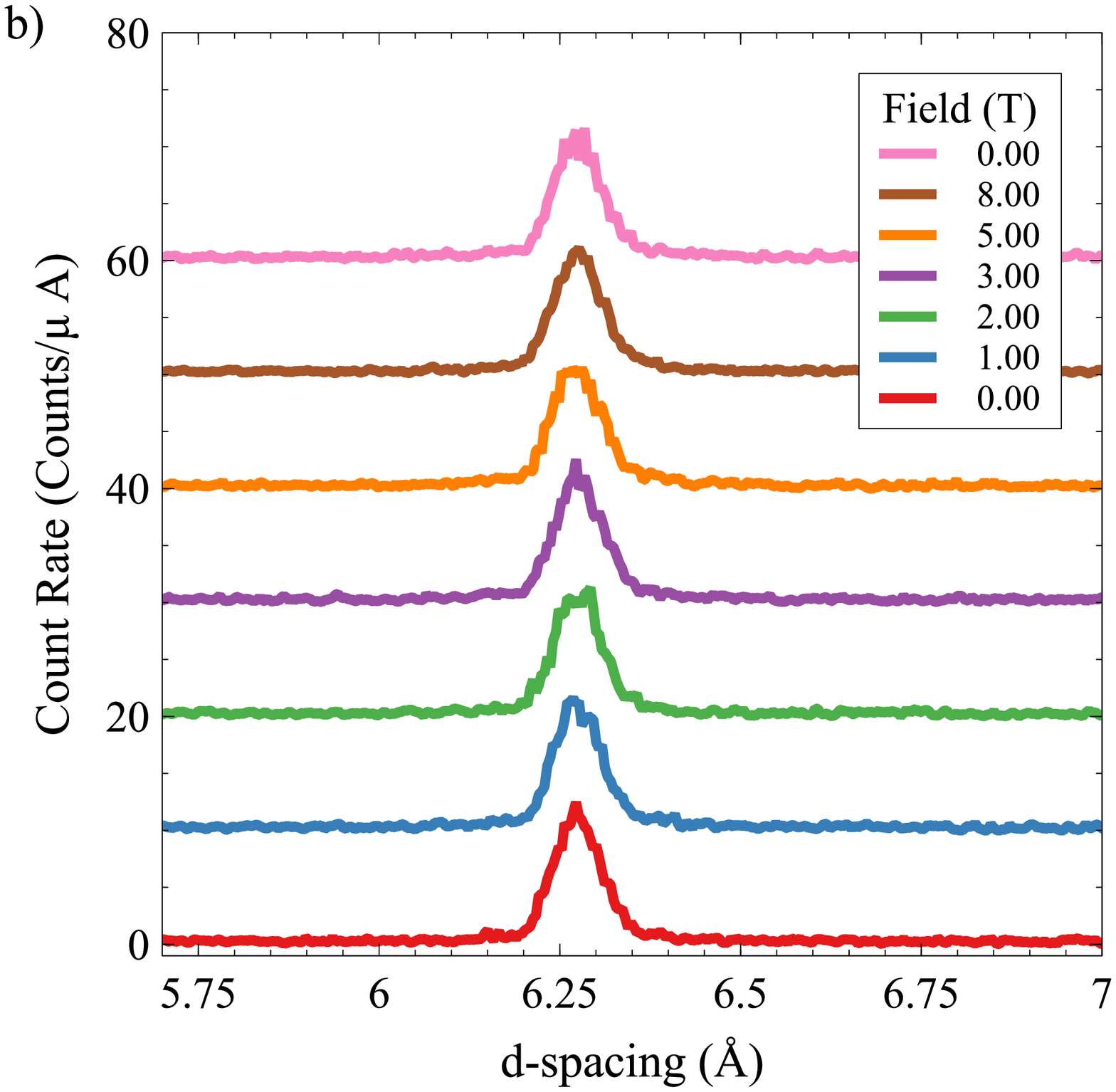}
\caption{\label{fig:rc194dwaterfalls} Waterfall plots of  a) in plane and b) out of plane peaks for the 45~nm film with increasing field applied along the hard axis. The peak amplitudes correspond to the blue diamonds in Figure S1 a,c.}
\end{figure}

\begin{figure}
\centering
\includegraphics[width=0.45\textwidth]{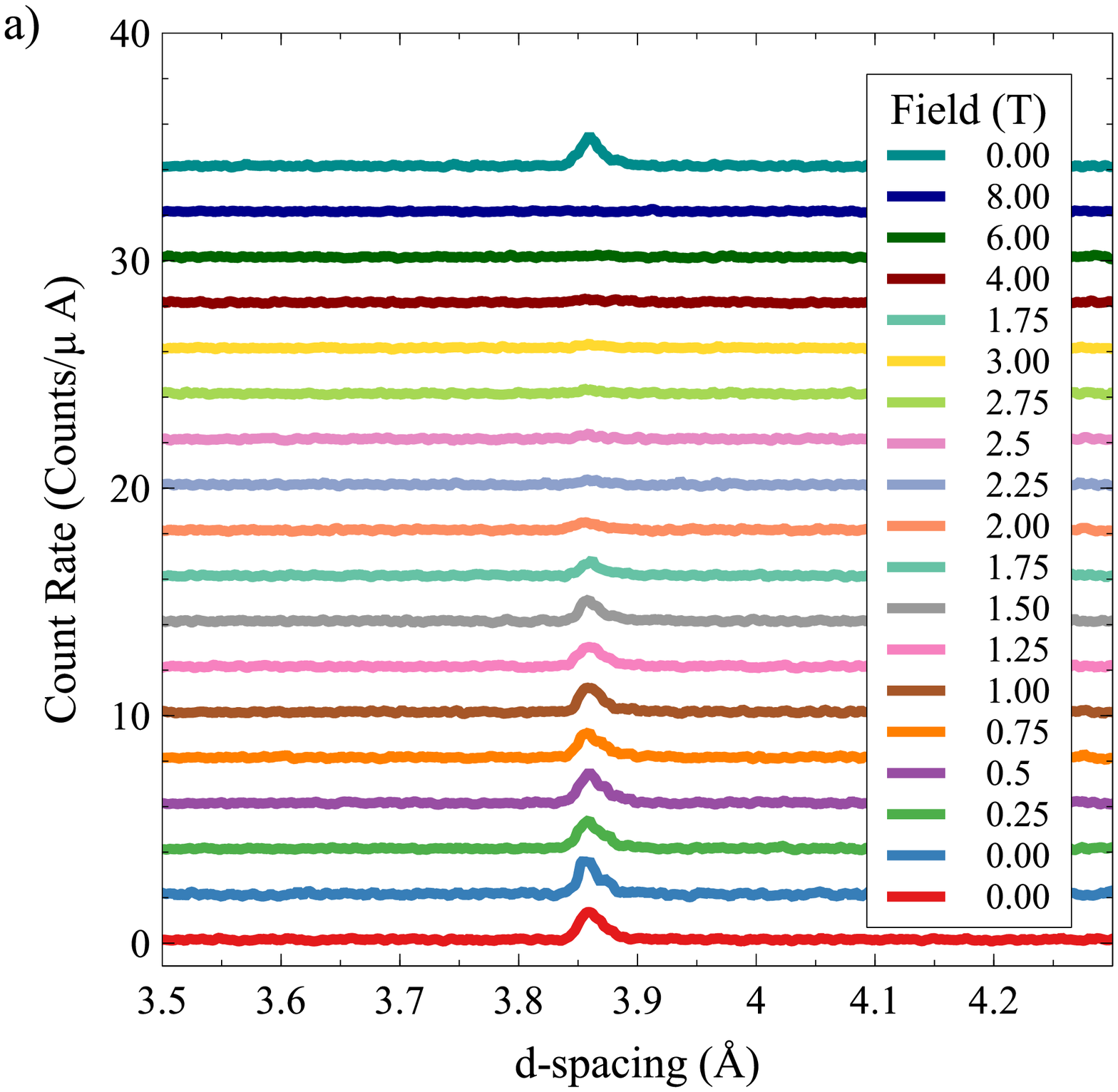}
\includegraphics[width=0.454\textwidth]{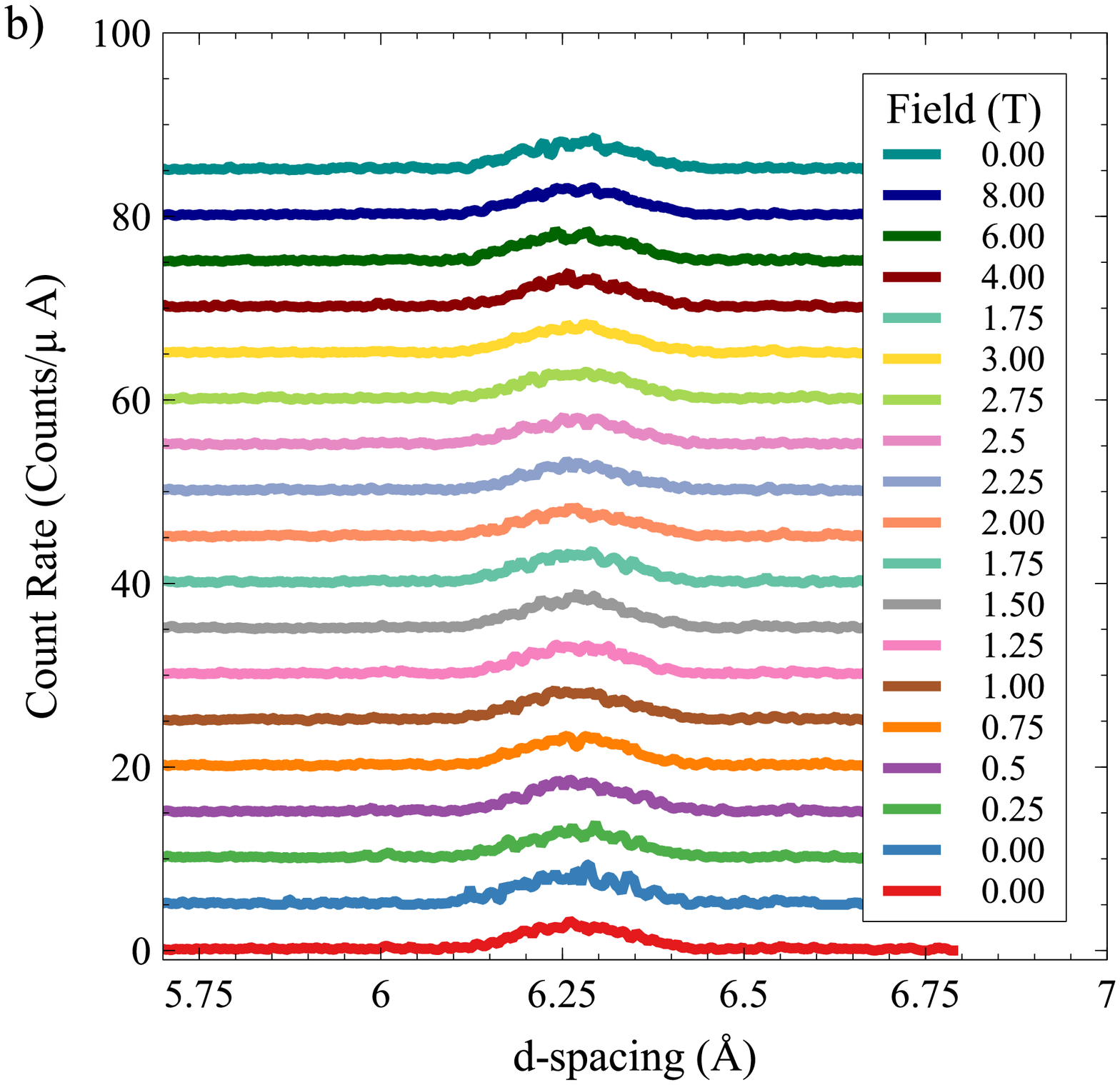}
\caption{\label{fig:rc123awaterfalls} Waterfall plots of a) in plane and b) out of plane peaks for the 20~nm film with increasing field applied along the easy axis. The peak amplitudes correspond to the red circles in Figure S1 b,d.}
\end{figure}

\begin{figure}
\centering
\includegraphics[width=0.45\textwidth]{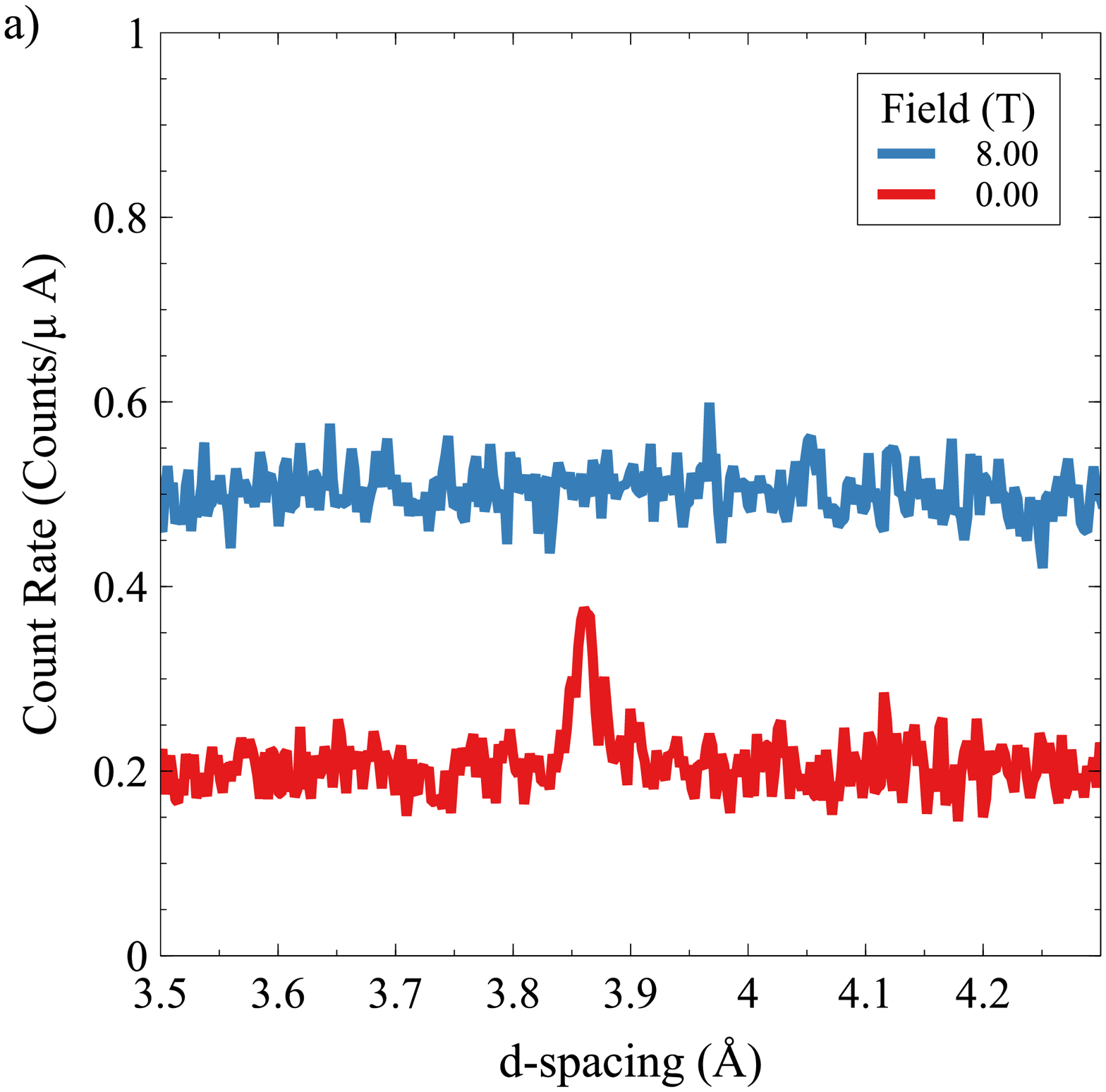}
\includegraphics[width=0.454\textwidth]{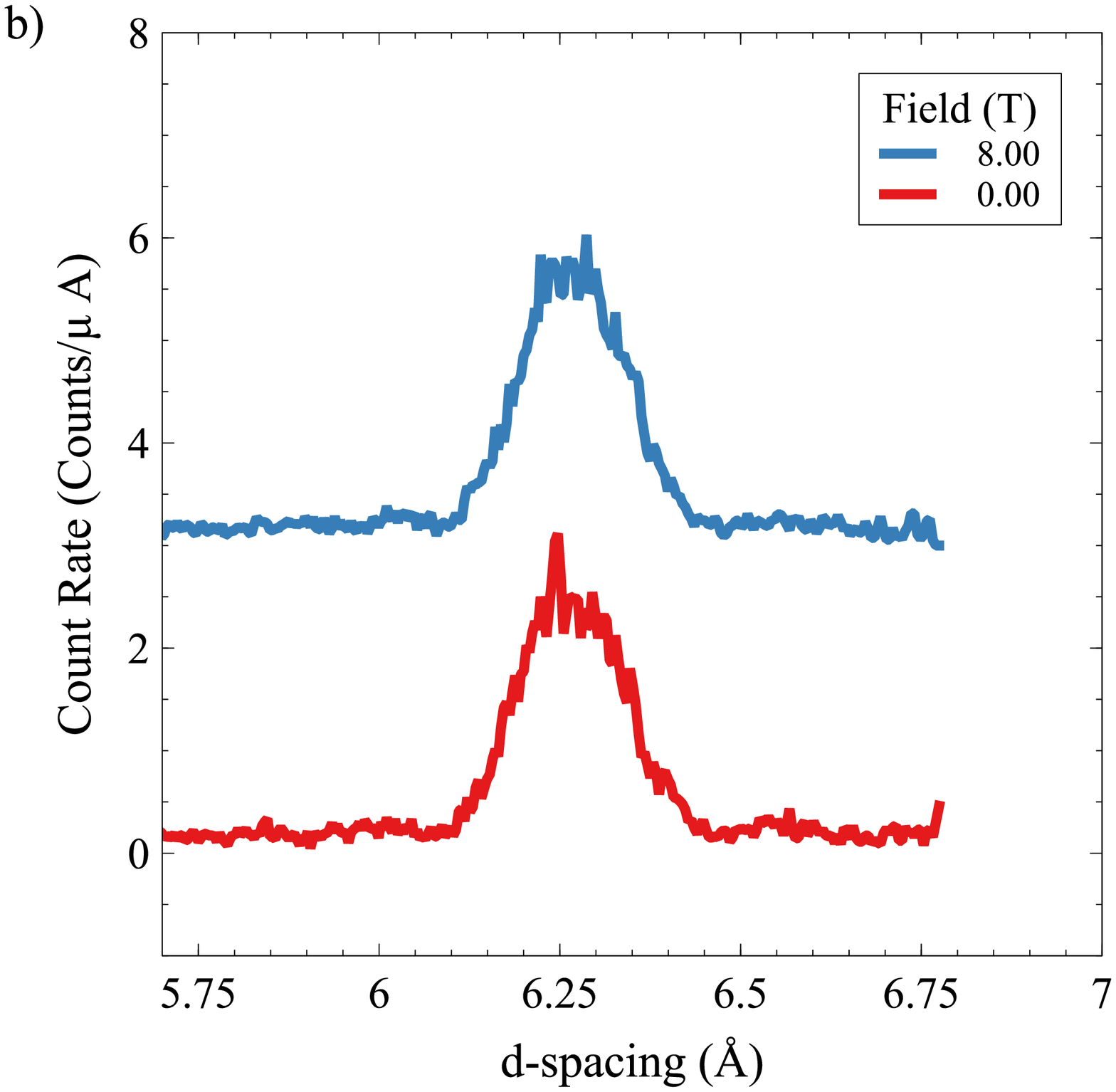}
\caption{\label{fig:rc123bwaterfalls} Waterfall plots of  a) in plane and b) out of plane peaks for the 20~nm film with increasing field applied along the hard axis. The peak amplitudes correspond to the blue diamonds in Figure S1 b,d.}
\end{figure}